\documentclass[12pt]{article}
\usepackage{graphicx}
\usepackage{epsfig}
\usepackage{amssymb,amsmath}
\usepackage{pstricks}

\input{epsf.sty}

\newcommand{\be}{\begin{equation}}
\newcommand{\ee}{\end{equation}}
\newcommand{\ba}{\begin{eqnarray}}
\newcommand{\ea}{\end{eqnarray}}
\newcommand{\bi}{\begin{itemize}}
\newcommand{\ei}{\end{itemize}}

\newcommand{\Tr}{{\rm Tr\,}}

\newcommand{\ex}{{\rm e}}

\newcommand{\nn}{\nonumber}

\newcommand{\bfx}{{\bf x}}
\newcommand{\bfy}{{\bf y}}

\newcommand{\RR}{{\rm I\kern -.2em  R}} 
\newcommand{\eq}{Eq.~}
\newcommand{\eqs}{Eqs.~}
\newcommand{\fig}{Fig.~}

\def\lsi{\raise0.3ex\hbox{$<$\kern-0.75em\raise-1.1ex\hbox{$\sim$}}}
\def\gsi{\raise0.3ex\hbox{$>$\kern-0.75em\raise-1.1ex\hbox{$\sim$}}}

\newcommand{\gsim}{\mathop{\gsi}}

\begin{document}
 
\begin{titlepage}
\begin{flushright}
MS-TP-08-11
\end{flushright}
\begin{centering}
\vfill
 
{\bf\Large Strong coupling expansion for finite temperature Yang-Mills theory in the 
confined phase }

\vspace{0.8cm}
 
Jens Langelage, Gernot M\"unster and Owe Philipsen

\vspace{0.3cm}
{\em 
Institut f\"ur Theoretische Physik, Westf\"alische Wilhelms-Universit\"at M\"unster, \\
48149 M\"unster, Germany}

\vspace*{0.7cm}
 
\begin{abstract}
We perform Euclidean strong coupling expansions for Yang Mills theory on the 
lattice at finite temperature. After setting up the formalism
for general SU(N), we compute the first few terms of the 
series for the free energy density and the lowest screening mass in the case of SU(2).
To next-to-leading order the free energy series agrees with that of an ideal gas of 
glueballs. This demonstrates that in the confined phase
the quasi-particles indeed correspond to the $T=0$ hadron excitations, as commonly assumed
in hadron resonance gas models.
Our result also fixes the lower integration constant for Monte
Carlo calculations of the thermodynamic pressure via the integral method.
In accord with Monte Carlo results, we find screening masses to be nearly temperature 
independent in the confined phase.
This and the exponential smallness of the pressure can be understood as genuine 
strong coupling effects. 
Finally, we analyse Pad\'e approximants to estimate the critical couplings of the 
phase transition, 
which for our short series are only $\sim 25\%$ accurate. However, up to these
couplings the equation of state agrees quantitatively with numerical results on $N_t=1-4$
lattices. 
\end{abstract}
\end{centering}
 
\noindent
\vfill
\noindent
 

\vfill

\end{titlepage}
 

\section{Introduction}

The study of QCD at finite temperatures and densities is of growing phenomenological interest for
current and future heavy ion collision experiments, as well as for many astrophysical
problems. Due to the interaction strength, the usual perturbative treatment by a series expansion
in a small coupling constant fails for QCD. 
Perturbative failure persists in the quark gluon plasma phase, where the 
weak coupling series is only defined up to a certain maximal order
(depending on the observable) before the well-known Linde infrared
problem sets in \cite{linde}. 
The best one can do is to calculate an effective
theory for the infrared modes with perturbatively calculable coefficients,
such as dimensional reduction \cite{dr} or hard thermal loops \cite{htl},
and solve the effective theory for the soft modes on the lattice.
The fully non-perturbative alternative are of course 
lattice Monte Carlo simulations. 
For a recent review of lattice results, see \cite{op}.

On the other hand, for temperatures below the plasma transition, 
no analytic approaches starting from the QCD Lagrangian are available at all.
Even lattice simulations of the equation of state are difficult in this regime, due
to the exponential suppression of the pressure with hadronic masses at low temperatures.  
Moreover, numerical determinations of the pressure via the integral method \cite{Boyd:1996bx} 
require to supply a lower integration constant, 
corresponding to the pressure at some low temperature.
Being unknown from first principles, this constant is usually set to zero
by hand, based on its assumed exponential smallness.
A successful description of lattice data below $T_c$ is given by the hadron resonance gas 
model \cite{hgm}, which requires expansive experimental knowledge of the hadron spectrum
as well as some modelling.

In the present work we fill these gaps with an analytic treatment of the low 
temperature phase by means of a strong coupling expansion of the lattice pure gauge theory.
Contrary to weak coupling expansions, strong coupling expansions are known to
be convergent series with a well-defined radius of convergence. 
In the early days of lattice gauge theory they were
used to get analytical results for some physical quantities of interest,
such as glueball masses \cite{sct0} or the energy density of lattice Yang-Mills 
theories \cite{balian,drouffe}.
These calculations were done at zero temperature, i.e.~for lattices with infinite
spatial volume $N_s^3$ and temporal extent $N_t$. To our knowledge, strong coupling 
expansions of the thermal partition function have not been considered beyond the 
infinite coupling limit, thus neglecting gauge fluctuations.
In the Euclidean framework, an effective theory for Wilson lines can be constructed in 
the strong coupling limit \cite{pol}, to be analysed by mean field methods. 
The only series we are aware of is for the temperature-dependent string tension \cite{green}.
In Hamiltonian approaches, an effective Hamiltonian for the strong coupling limit is constructed, 
and the partition function still has to be solved for by some other means. For the pure gauge
theory, a deconfinement transition was predicted in this way \cite{ham1}.
More recent applications are to systems with fermions at finite density, e.g.~\cite{ham}.
For a review of early work and references, see \cite{ben}. 

In this work, we calculate Euclidean strong coupling series for the free energy density
and screening masses in SU(2) pure gauge theory with an infinite
spatial volume and a compactified temporal lattice extent $N_t$.
In this way we can analytically study finite temperature effects in the confined phase.
The physical deconfinement phase transition
then corresponds to a finite convergence radius of the series, which we try to estimate
from the behaviour of the coefficients. In Sec.~\ref{form} we set up the formalism of
computing strong coupling series for the free energy at finite temperature. Sec.~\ref{su2}
gives the explicit series for the SU(2) gauge theory and discusses how to 
leading order it coincides with that of a non-interacting glueball gas, as well as 
estimates for the radius of convergence and the phase transition. We also compare
our results to Monte Carlo simulations. Sec.~\ref{mass} discusses strong coupling series
of screening masses before we conclude in Sec.~\ref{con}.

\section{The free energy density as strong coupling series \label{form}}

Consider SU(N) Yang-Mills theory on an $\Omega=N_s^3\times N_t$ 
lattice with lattice spacing $a$ and the Wilson action. Its partition function is given by 
\ba
Z&&=\int DU \, \ex^{-S(U)}=\int DU\,\prod_p\ex^{-S_p(U)}\nn\\&&=\int DU\,\exp
\sum_{x}\sum_{1\leq\mu<\nu\leq 4}\beta\left(1-\frac{1}{N}{\rm Re}
\Tr U_p(x)\right), \quad
\beta=\frac{2N}{g^2}.
\label{defz}
\ea
Here, $g^2$ is the coupling constant of the corresponding 
continuum field theory, and the elementary 
plaquettes are given in terms of link variables 
as $U_p(x)=U_\mu(x)U_\nu(x+a\hat{\mu})U^\dag_\mu(x+a\hat{\nu})U^\dag_\nu(x)$.
The finite space-time box corresponds to a physical volume $V=(aN_s)^3$ and a temperature
$T=1/(aN_t)$. We are interested in the thermodynamic limit and always consider
the situation $N_s\rightarrow\infty$. 

To our knowledge, Euclidean strong coupling expansions 
have so far only been applied to the $T=0$ or $N_t\rightarrow\infty$ situation.
Formally, the partition function \eq(\ref{defz}) still resembles a thermal system 
in this limit,
with the lattice gauge coupling $\beta$ playing the role of inverse temperature.
Correspondingly, strong coupling expansions in small $\beta$ are often termed
`high temperature expansions' in the literature \cite{mm}, 
even though the physical temperature of the system is zero.
We shall not repeat the derivation of the strong coupling expansion here, but only give
some central formulae required to fix the notation.
For more details, see \cite{mm} and references therein.
To start with, the exponential of the action is expanded in group characters 
$\chi_r(U)=\Tr D_r(U)$, with $D_r(U)$ a specific irreducible representation matrix of $U$ with dimension
$d_r$,
\be
\ex^{-S_p(U)}=c_0(\beta)\left[1+\sum_{r\neq 0}d_ra_r(\beta)\chi_r(U)\right],\quad
a_r=\frac{c_r(\beta)}{c_0(\beta)}. 
\ee
The coefficients of the character expansion $c_r(\beta)$, 
and hence the effective expansion parameters 
$a_r(\beta)$, can be expanded in powers of $\beta$ to yield the desired series in $g^{-2}$.
The series can then be reorganised as a sum of graphs $G$ with contributions $\Phi(G)$,
\be
Z=c_0^{6\Omega}\sum_G \Phi(G),\qquad \Phi(G)=\int DU\prod_{p\in G} d_{r_p} a_{r_p} \chi_{r_p}(U)
=\prod_i \Phi(X_i),
\ee
which factorise into disconnected components $X_i$, called polymers.
Finally, using the formalism of moments and cumulants, one arrives at a cluster
expansion for our quantity of interest, the free energy density,
\begin{equation}
\tilde{f}\equiv-\frac{1}{\Omega}\ln Z=-6\ln\,c_0(\beta)-\frac{1}{\Omega}
\sum_{C=(X_i^{n_i})}\,a(C)\prod_i\Phi(X_i)^{n_i}.
\label{free}
\end{equation}
The sum is over all clusters $C$, which are defined as connected polymers $X_i$, and 
$n_i$ denotes the multiplicity of a particular polymer in a cluster.
The combinatorial factor $a(C)$ is given as 
\begin{equation}
a(C)=\dfrac{[X_1,\dots,X_1,X_2,\dots,X_2,\dots,X_k]}{n_1!n_2!\dots n_k!}
\end{equation}
and equals $1$ for clusters $C$ which consist of only one polymer $X_i$. The so-called cumulant $[$ $]$ can be expressed in terms of moments $<$ $>$
\begin{equation} 
[\alpha,\dots,\zeta]=\sum_P(-1)^{n-1}(n-1)!<\alpha,\dots,\beta>\dots<\gamma,\dots,\delta>
\end{equation}
where $n$ is the number of factors on the right hand side and the sum goes over all partitions $P$. The moments are defined in such a way, that 
\begin{equation}
<X_1,\dots,X_n>=\left\lbrace \begin{array}{l} 1,\mbox{ if every pair }X_i,X_j\mbox{ is disconnected} \\ 0,\mbox{ otherwise} \end{array}\right.
\end{equation}
This implies the equivalence between non-zero cumulants and connectedness of graphs,
\begin{equation}
 [X_1,\dots,X_n]\neq0\quad\Leftrightarrow\quad X_1\cup\dots\cup X_n\quad\mbox{ is connected.}
\end{equation}
The contributing polymers $X_i$ have to be objects with a closed surface since
\begin{eqnarray}
\int dU \chi_r(U)=\delta_{r,0}.
\label{trivial}
\end{eqnarray}
This means the group integration projects out the trivial representation at each link. 
Group integrals are calculated using integration formulae like
\begin{equation}
\int dU \chi_r(UV)\chi_s(WU^{-1})=\frac{\delta_{rs}}{d_r}\chi_r(VW).
\end{equation}
Note that, because of translation invariance and connectedness, the number of identical clusters at different
positions is $\propto \Omega$, so that the lattice volume drops out of \eq (\ref{free}) which
thus has a finite thermodynamic limit. 

We now wish to apply this formalism to the case of non-zero {\it physical}
temperature, which is realized by keeping $N_t=1/(aT)$ finite.
The free energy density related to a physical temperature $T$ is defined as
\be
\frac{F}{V}=- \frac{T}{V}\ln Z = -\frac{1}{N_tN_s^3}\ln Z=\tilde{f}(N_t),
\label{fphys}
\ee
i.e.~it corresponds to the previously defined free energy density evaluated at finite $N_t$.
Evidently, all formulae above remain unchanged in this case. 
The only effect of finite $N_t$ with periodic boundary conditions
is to change the set of contributing graphs $\{X_i\}$.
The physical free energy is then obtained by subtracting the formal 
free energy $\tilde{f}(N_t=\infty)$, which renormalises \eq(\ref{fphys}) 
analogous to a subtraction of the divergent vacuum energy in the continuum,
\begin{equation}
f(N_t,\beta)=\tilde{f}(N_t,\beta)-\tilde{f}(\infty,\beta).
\label{freephys}
\end{equation}

\subsection{Classification of graphs \label{graphs}}

Because of the difference in \eq(\ref{freephys}), those graphs contributing in the same way to 
$\tilde{f}(N_t)$ and $\tilde{f}(\infty)$ drop out of the physical free energy. This is true for
all polymers with time extent less than $N_t$.  
The calculation thus reduces to graphs with a temporal size of $N_t$ on the 
finite $N_t$ lattice, and graphs spanning or extending $N_t$ on the infinite lattice.
Such graphs contribute either to $\tilde{f}(N_t)$ or to $\tilde{f}(\infty)$ 
(and in some cases to both), and hence
to the difference in \eq\ref{freephys}. 
It is therefore clear from the outset that 
the strong coupling series for the physical free energy starts at a higher
order than the formal zero temperature free energy. Moreover, the order of the leading 
contribution depends on $N_t$. 

\begin{figure}
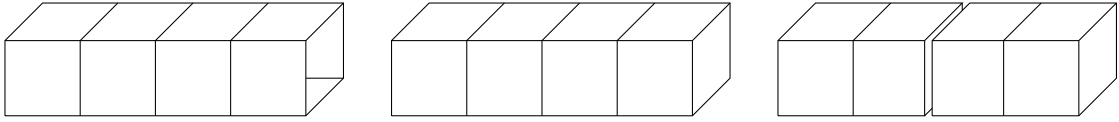

\begin{center}
\vspace*{2cm}
\begin{minipage}{5cm}
\scalebox{0.5}{
\psline(0,0)(8,0)(9,1)(9,3)(1,3)(0,2)(0,0)
\psline(0,2)(8,2)(9,3)
\psline(2,0)(2,2)(3,3)
\psline(4,0)(4,2)(5,3)
\psline(6,0)(6,2)(7,3)
\psline(8,0)(8,2)
\psline(9,1)(8,1)
}
\end{minipage}
\begin{minipage}{5cm}
\scalebox{0.5}{
\psline(0,0)(8,0)(9,1)(9,3)(1,3)(0,2)(0,0)
\psline(0,2)(8,2)(9,3)
\psline(2,0)(2,2)(3,3)
\psline(4,0)(4,2)(5,3)
\psline(6,0)(6,2)(7,3)
\psline(8,0)(8,2)
}
\end{minipage}
\begin{minipage}{5cm}
\scalebox{0.5}{
\psline(0,0)(3.9,0)(3.9,2)(4.9,3)(1,3)(0,2)(0,0)
\psline(2,0)(2,2)(3,3)
\psline(0,2)(3.9,2)
\psline(4.9,3)(4.9,2.8)
\psline(3.9,0)(4.1,0.2)
\psline(4.1,0)(8,0)(9,1)(9,3)(5.1,3)(4.1,2)(4.1,0)
\psline(6,0)(6,2)(7,3)
\psline(8,0)(8,2)(9,3)
\psline(4.1,2)(8,2)
}
\end{minipage}
\caption{Graphs that appear on the different lattices. Left: Leading order tube for $N_t=4$. 
Middle: First graph vanishing in the case $N_t=4$. 
Right: Graph contributing on both, $N_t=4,\infty$, lattices.}\label{tubes}
\end{center}
\end{figure}

The lowest order graph existing due to the boundary condition on the finite 
$N_t$ lattice, but not on the infinite lattice, 
is a tube of length $N_t$ with a cross-section of one single plaquette, 
as shown in \fig\ref{tubes} (left). 
It forms a closed torus through the periodic boundary and thus gives a non-vanishing 
contribution, which is easily calculated to be $\Phi(G_1)=a_f^{4N_t}$, where the subscript '$f$'
indicates the fundamental representation.
We need to sum up all such graphs on the lattice. There are three spatial directions
for the cross section of the tube, giving a factor of 3. Translations in time take the graph into 
itself and do not give a new contribution, while we get $V\Phi(G_1)$ from all spatial 
translations. Together with the $1/\Omega$ in \eq\ref{free} this gives a factor of $1/N_t$.
The contribution of all tubes with all plaquettes in the fundamental representation is thus 
\be
\Phi(G_1)=\frac{3}{N_t}a_f^{4N_t},
\ee
which is - up to a sign - also the leading order result for the physical free energy. 
For SU(N) with $N\geq3$ we have an additional factor of 2 because there are also complex 
conjugate fundamental representations.

On the other hand, the same tube with both ends closed off by additional plaquettes as in 
\fig\ref{tubes} (middle), contributes with $\Phi(G_2)=d_f^2a_f^{4N_t+2}$ on an $N_t=\infty$, 
but not on a finite $N_t$ lattice. This is because
the boundary plaquettes get identified as one doubly occupied plaquette, 
which is not an allowed graph in the expansion. Therefore,
$\Phi(G_2)$ counts with a negative sign relative to $\Phi(G_1)$ towards the physical free energy. 
Here, translations in time do produce a new graph, so the total contribution is
\be
\Phi(G_2)=-3d_f^2a_f^{4N_t+2}.
\ee
 
\fig\ref{tubes} (right) shows a variation of these basic graphs, 
a cluster composed of two double cubes.
This is an example of a graph spanning $N_t$ which contributes to both 
the finite and infinite $N_t$ lattices in the same way, thus cancelling out in the 
physical free energy. (Note that the corresponding
tube obtained without the plaquettes at $N_t$ vanishes, it would correspond to a single 
polymer with doubly occupied plaquettes at the slit). However, for similar 
clusters composed of more than two polymers, this cancellation in general no longer holds
because of different assignments of combinatoric
factors $a(C)$ in the two cases.

\subsection{Corrections to basic polymers}

\begin{figure}
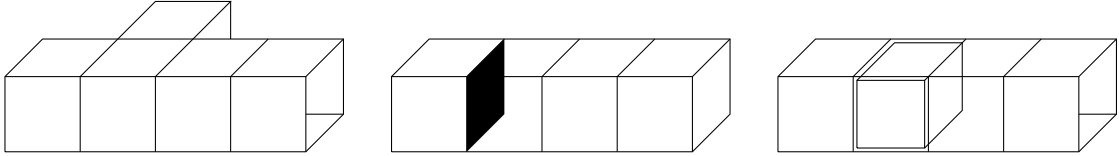

\begin{center}
\vspace*{2cm}
\begin{minipage}{5cm}
\scalebox{0.5}{
\psline(0,0)(8,0)(9,1)(9,3)(1,3)(0,2)(0,0)
\psline(0,2)(8,2)(9,3)
\psline(2,0)(2,2)(3,3)
\psline(4,0)(4,2)(5,3)
\psline(6,0)(6,2)(7,3)
\psline(8,0)(8,2)
\psline(9,1)(8,1)
\psline(3,3)(4,4)(6,4)(5,3)
\psline(6,4)(6,3)
}
\end{minipage}
\begin{minipage}{5cm}
\scalebox{0.5}{
\psline(0,0)(8,0)(9,1)(9,3)(1,3)(0,2)(0,0)
\psline(0,2)(8,2)(9,3)
\psline(2,0)(2,2)(3,3)
\psline(4,0)(4,2)(5,3)
\psline(6,0)(6,2)(7,3)
\psline(8,0)(8,2)
\pspolygon*[fillcolor=black](2,0)(3,1)(3,3)(2,2)(2,0)
}
\end{minipage}
\begin{minipage}{5cm}
\scalebox{0.5}{
\psline(0,0)(8,0)(9,1)(9,3)(1,3)(0,2)(0,0)
\psline(0,2)(8,2)(9,3)
\psline(2,0)(2,2)(3,3)
\psline(4,0)(4,2)(5,3)
\psline(6,0)(6,2)(7,3)
\psline(8,0)(8,2)
\psline(9,1)(8,1)
\psline(2.1,0.1)(3.9,0.1)(3.9,1.9)(2.1,1.9)(2.1,0.1)
\psline(2.1,1.9)(3.1,2.9)(4.9,2.9)(4.9,1.1)(3.9,0.1)
\psline(3.9,1.9)(4,2)
}
\end{minipage}
\end{center}
\caption{Graphs contributing to the higher order terms of the series.}\label{tubes3}
\end{figure}

For fixed $N_t$, starting from the basic leading order polymers discussed in the 
last section, 
one can now build up the corrections by adding decorations on each of them. 
These can be either geometric, by adding additional 
fundamental representation plaquettes as in \fig\ref{tubes3} (left), 
 by inserting plaquettes in 
higher representations as in \fig\ref{tubes3} (middle) or by adding a whole 
new polymer as in \fig\ref{tubes3} (right). Of course, these modifications can be combined.
Adding plaquettes in higher representations is possible only if at each and every link the 
Clebsch-Gordan series of the representation matrices contains the trivial representation, 
due to \eq(\ref{trivial}).

We have already seen in Sec.~\ref{graphs} that the order in $a_f$, to which the 
graphs contribute, depends on $N_t$. Thus, the relative importance of different
types of graphs changes with $N_t$. For our example $N_t=4$ considered in Sec.~\ref{graphs},
the leading correction to the basic polymer on the $N_t$ lattice is the insertion of one
higher representation plaquette inside the tube, contributing an additional factor 
$\sim a_f^2$, whereas
the lowest order geometric decoration is a shifted plaquette adding a factor $\sim a_f^4$.
By contrast, on an $N_t=1$ lattice contributions to the leading order correction $\sim a_f^2$ 
come from the basic polymer with a cross sectional perimeter of six links. 
In general, geometric decorations enter
earlier the lower $N_t$ is. Thus the summation of basic polymers and their 
decorations contains the complete result to some fixed 
order $O$ only for lattices $N_t\geq N_t^O$, with some $N_t^O$ which is obviously 
growing with $O$. For lattices $N_t<N_t^O$ there are additional geometric decorations
contributing to $O(a_f^O)$. In this work, we have calculated corrections to 
$a_f^{4N_t}$ through $O(a_f^8)$ for which $N_t^O=5$.

\subsection{Summing basic polymers and their corrections for SU(2)}

The contribution of all graphs of length $N_t$ without geometric decorations can 
be summed up in closed form. To do this we note 
that an additional plaquette in a representation $r$ as in \fig\ref{tubes3} (middle) 
gives an additional factor of $d_ra_r$. In the case of SU(2), for which we will describe the calculation, the only possibility is the $j=1$ representation, because of \eq(\ref{trivial}) and
\begin{equation}
\frac{1}{2}\otimes\frac{1}{2}=0\oplus 1\quad\Rightarrow\quad\frac{1}{2}\otimes\frac{1}{2}\otimes 1=0\oplus1\oplus1\oplus2 \;.
\end{equation}
The expansion parameters of the lowest representations are given by modified Bessel functions
and can be expanded in powers of the lattice coupling,
\begin{eqnarray}
u&\equiv&a_{1/2}=\dfrac{I_2(\beta)}{I_1(\beta)}=\frac{1}{4}\beta-\frac{1}{96}\beta^3+\frac{1}{1536}\beta^5-\frac{1}{23040}\beta^7+O(\beta^9),\nn\\
v&\equiv&a_{1}=\dfrac{I_3(\beta)}{I_1(\beta)}=
\frac{2}{3}\,{u}^{2}+\frac{2}{9}\,{u}^{4}+{\frac {16}{135}}\,{u}^{6}+{\frac {8}{135}}\,{u}^{8}+{\cal{O}}(u^{10}).
\end{eqnarray}
For the following it is convenient to introduce the combination $c=1+3v-4u^2$ and to use $u=a_f$ 
as the effective expansion parameter instead of $\beta$.
It is well known from expansions at zero temperature that apparent convergence
is better for the series in $u$ \cite{sct0,drouffe}, and we observe the same phenomenon here.

On the $N_t$ lattice we can have $0\leq k\leq N_t$ additional plaquettes at $N_t$ places which can be distributed in $\binom{N_t}{k}$ ways. Summing over all possible distributions gives
\begin{equation}
\Phi_1=\Phi(G_1)\sum_{k=0}^{N_t}\binom{N_t}{k}\left(3v\right)^k=\Phi(G_1)\left(1+3v\right)^{N_t}.
\end{equation}

We can also add slits to get graphs as in \fig\ref{tubes} (right), consisting of more than one polymer. 
Each slit gives a factor $d_f^2u^2=4u^2$. The minimum number of slits, $i$, 
is 2 and the combinatorial factor for such graphs is
\begin{equation}
 a(C)=(i-1)(-1)^{i-1}.
\end{equation}
Summing these graphs with possible $j=1$ plaquettes at the remaining places we get
\begin{eqnarray}
 \Phi_2&=&\Phi(G_1)\sum_{i=2}^{N_t}\binom{N_t}{i}(i-1)(-1)^{i-1}\left(4u^2\right)^{i}\sum_{k=0}^{N_t-i}\binom{N_t-i}{k}\left(3v\right)^k\nonumber\\
&=&\Phi(G_1)\left[c^{N_t}-\left(1+3v\right)^{N_t}+4u^2N_tc^{N_t-1}\right].
\end{eqnarray}

Of course, we can make the same insertions to the graph \fig\ref{tubes} (middle) on the infinite lattice with the difference that we have at most $N_t-1$ places to add plaquettes. The combinatorial factor now reads
\begin{equation}
 a(C)=(-1)^{i},
\end{equation}
where $i$ is again the number of slits and, in this case, it is unrestricted, giving 
\begin{eqnarray}
\Phi_3&=&\Phi(G_2)\sum_{i=0}^{N_t-1}\binom{N_t-1}{i}(-1)^{i}\left(4u^2\right)^i
\sum_{k=0}^{N_t-1-i}\binom{N_t-1-i}{k}\left(3v\right)^k\nonumber\\
&=&\Phi(G_2)c^{N_t-1}\nonumber\\
&=&\Phi(G_1)(-4u^2N_t)c^{N_t-1}.
\end{eqnarray}
For the final result we have to add the different pieces and get
\be
 \Phi=\Phi_1+\Phi_2+\Phi_3
=\dfrac{3}{N_t}u^{4N_t}c^{N_t}.
\ee
In higher gauge groups, the summation proceeds in a similar fashion with some slight 
modification due to the fact that there are also complex conjugate representations.

\section{The free energy density for SU(2) \label{su2}}
 
Let us now give our central results for the gauge group SU(2). 
For $N_t\geq 5$, the basic polymers and their decorations
can be summed up to give the general result
\be
f(N_t,u)=-\frac{3}{N_t}u^{4N_t}c^{N_t}\left[1+12N_tu^4
-\frac{1556}{81}N_tu^6+\left(83N_t^2+\frac{41417}{243}N_t\right)u^8+O(u^{10})
\right].
\label{nt5}
\ee
For $N_t=1-4$ there are additional geometric decorations,
while some graphs contained in the previous result
do not contribute on those short lattices. We find
\begin{eqnarray}
f(1,u)&=&-3\,{u}^{4}-16\,{u}^{6}-{\frac {10913}{54}}{u}^{8}
-{\frac {968642}{405}}{u}^{10}+O \left( {u}^{12} \right)\\
f(2,u)&=&-\frac {3}{2}{u}^{8}+6\,{u}^{10}-55\,{u}^{12}+{\frac {29236}{135}}{u
}^{14}-{\frac {78413341}{43740}}{u}^{16}+O \left( {u}^{18} \right)\label{nt2}\\
f(3,u)&=&-{u}^{12}+6\,{u}^{14}-50\,{u}^{16}+{\frac {37966}{135}}{u}^{18}-{
\frac {856048}{405}}{u}^{20}+O \left( {u}^{22} \right)\label{nt3}\\
f(4,u)&=&-{\frac {3}{4}}{u}^{16}+6\,{u}^{18}-56\,{u}^{20}+{\frac {51376}{135}}{
u}^{22}-{\frac {2402453}{810}}{u}^{24}+O \left( {u}^{26} \right). \label{nt4}
\end{eqnarray}
The fact that only even powers of $u$ appear is due to the reality of the  
SU(2) representations. Note that for $N_t=1$, our series is shorter since we did not compute
the numerous geometric decorations at the next order.
From these expressions all other thermodynamic quantities of interest can be constructed, in particular
the pressure and energy density, respectively,
\be
p=-f,\quad 
e(\beta)=\frac{1}{6}\frac{d}{d\beta}f(\beta)=\frac{1}{6}\frac{du}{d\beta}\frac{d}{du}f(u).
\ee
Since the partition function is not directly measurable in Monte-Carlo simulations,
the pressure is usually obtained by the integral method \cite{Boyd:1996bx}, where
the expectation values of derivatives are computed and then integrated numerically,
\begin{eqnarray*}
 \frac{p}{T^4}\,\bigg\vert_{\beta_0}^{\beta}=N_t^4\int_{\beta_0}^{\beta}d\beta' \left[ 6\langle\,{\rm Tr}\,U_p^0\,\rangle-3\langle\,{\rm Tr}\,U_p^t+{\rm Tr}\,U_p^s\,\rangle\right],
\end{eqnarray*}
where $\langle\,{\rm Tr}\,U_p^0\,\rangle$ denotes the plaquette expectation 
value on symmetric ($T=0$)lattices, $N_t=N_s\rightarrow\infty$, and 
$\langle\,{\rm Tr}\,U_p^{t,s}\,\rangle$ are those of space-time 
and space-space plaquettes for $N_t<N_s$.
The lower integration limit is usually set to zero by hand, arguing with an exponentially small pressure in the low temperature regime. Our results justify this assumption from first principles and allow to 
fix that value if desired.

\subsection{The free energy density from an ideal glueball gas}

In weak coupling expansions of the pressure, the leading term is the well-known 
Stefan-Boltzmann limit, describing a non-interacting gas of the constituent particles.
It is now interesting to ask how the QCD pressure can be interpreted in the strong 
coupling regime.
From the Wilson action it is clear that the strong coupling limit is also non-interacting.
However, as we have  noted already, in this limit the pressure is zero. 
Considering strong but finite couplings, and
recalling the first orders of the $T=0$ glueball mass calculations for SU(2) \cite{sct0}, 
\begin{eqnarray}
m(A_1^{++})&=&-4\ln\,u+2u^2-\frac{98}{3}u^4-\frac{20984}{405}u^6-\frac{151496}{243}u^8,\nn\\
m(E^{++})&=&-4\ln\,u+2u^2-\frac{26}{3}u^4+\frac{13036}{405}u^6-\frac{28052}{243}u^8,
\end{eqnarray}
we observe that through second order in the expansion the free energy can be written as
\be
f(N_t,u)=-\frac{1}{N_t}\left[e^{-m(A_1^{++})N_t} + 2e^{-m(E^{++})N_t} 
+ {\cal{O}}(u^4)\right]. 
\ee
Here, the prefactors 1 and 2 before the exponentials correspond to the number of 
polarisations of the respective glueball states. Note that higher spin states 
start with $\sim 6\ln u$ \cite{schor}, thus 
contributing to the order $\sim u^{6N_t}$ or higher in the free energy.
Hence, through two non-trivial orders our result
is that of a free glueball gas, modified by higher order corrections.
This is a rather remarkable result. 
It allows to see from a first principle calculation that the pressure is 
exponentially small in the confined phase, and that it is well approximated by an 
ideal gas of quasi-particles which correspond to
the $T=0$ hadron excitations. While this result might be expected on phenomenological
grounds, it is nice to see
it demonstrated by an explicit calculation.
This gives a quantum field theoretical explanation for the otherwise plausible 
success of the hadron-resonance-gas model in reproducing the confined phase equation of 
state \cite{hgm}.

\subsection{Series analysis and phase transition}

Strong coupling or high temperature expansions have been worked out to high orders in many
spin models, where various tools of series analysis can be applied to improve convergence or 
extract additional information from the behaviour of the coefficients \cite{drouffe, gutt}.
In this section we explore some of these possibilities with our series.
In particular, we are interested in the radius of convergence of the strong coupling series.
SU(N) pure gauge theories at finite temperature have a true order parameter for confinement,
the Polyakov loop, and therefore there is a non-analytic phase transition separating the
confined from the deconfined phase. The associated critical coupling $\beta_c$ limits
the radius of convergence of the strong coupling series, provided there are no other 
singularities
$\beta_s$ in the complex $\beta$-plane with $|\beta_s|<\beta_c$.  

\begin{figure}[t]
\hspace*{-0.7cm}
\includegraphics*[angle=-90,width=0.55\textwidth]{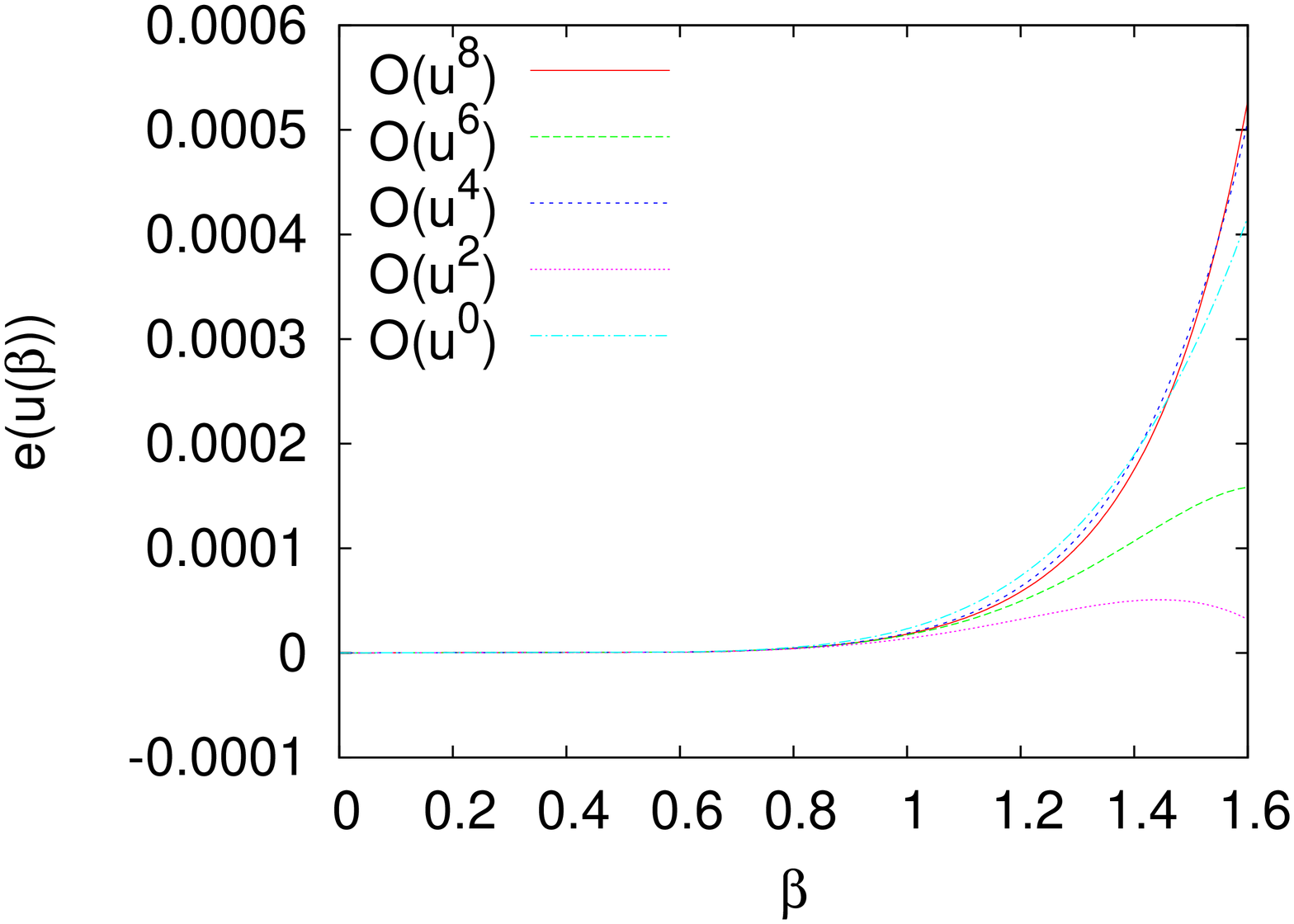}
\includegraphics*[angle=-90,width=0.55\textwidth]{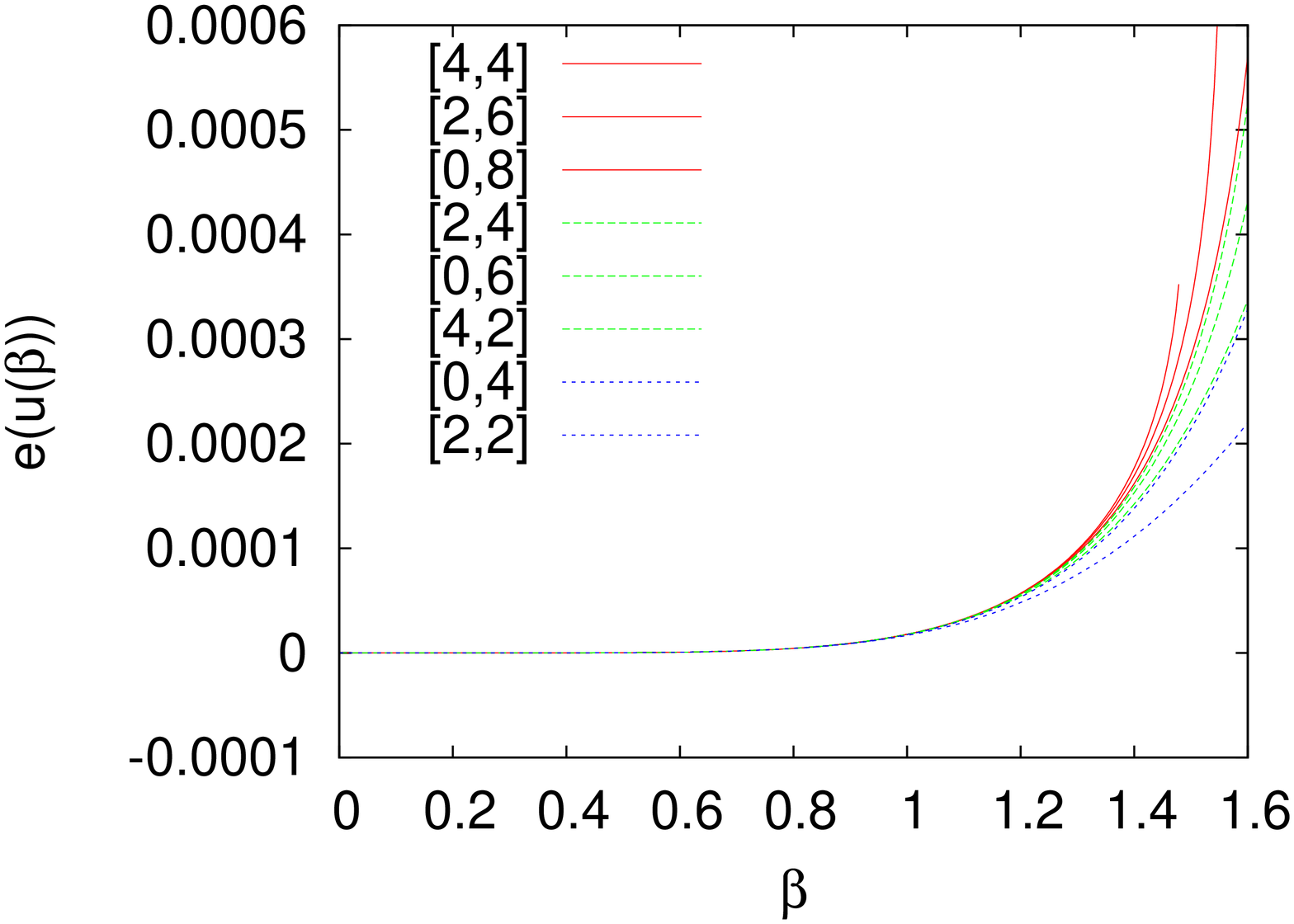}
\caption{The energy density in the confined phase on $N_t=2$. Left: subsequent orders of 
the strong coupling series. Right: Reconstructed from the $L+M=4,6,8$ Pad\'e approximants
to the strong coupling series of $D_C(u)$, \eqs(\ref{dc}),(\ref{conv}).
}
\label{series}
\end{figure}

For our later comparison with Monte Carlo results, 
it is particularly convenient to consider the energy density.
The curves for consecutive orders in the strong coupling expansion
for $N_t=2$ are plotted in \fig\ref{series} (left). 
For $\beta\gsim 1$ convergence rapidly becomes poor, announcing the proximity of the
convergence radius.
The most straightforward way to estimate the radius of convergence $r_n$ from the $n$th order
series would be via the ratio test of the series coefficients $f_{2n}$,
\be
r_n=\left|\frac{f_{2n}}{f_{2n+2}}\right|^{1/2}.
\ee 
However, our series \eqs(\ref{nt5}-\ref{nt4}) are still rather short and no convergence in
$r_n$ is visible. Moreover, a singularity on the real axis requires the coefficients 
asymptotically to come with equal signs. This is clearly not the case in our expressions for $N_t>1$,
which suggest a nearby imaginary singularity.

A much better tool for our purposes is the analysis of Pad\'e approximants to a function
constructed from its series expansion. These are the rational functions
\begin{eqnarray*}[L,M](u)\equiv 
\frac{a_0+a_1u+\dots +a_Lu^L}{1+b_1u+\dots+b_Mu^M},
\end{eqnarray*}
with coefficients $a_i,b_i$ chosen such that they reproduce the power series
of the function of interest to the degree $L+M$. 
As rational functions, Pad\'e approximants are known to give good estimates of isolated pole 
singularities, whereas branch cuts or algebraic singularities are less well reproduced 
(for a detailed discussion, see \cite{gutt}). Furthermore, they give access to
several singularities, rather than just the nearest one. At the finite temperature 
phase transition the free energy $f(N_t,u_c)$ with $u_c=u(\beta_c)$ is continuous, with
a discontinuous first or second derivative, depending on the order of the transition. 
This type of
singularity is difficult to model for Pad\'e approximants.
Instead, the `heat capacity' 
\be
C(N_t,u)=u^2\dfrac{d^2}{du^2}f(N_t,u)
\ee
diverges at the phase transition as $C(u)\sim (u_c-u)^\alpha$ with a critical exponent
characteristic of the transition. Its logarithmic derivative 
\be
D_C(N_t,u)\equiv\frac{d}{du}\ln C(N_t,u)\sim -\frac{\alpha}{u_c-u} 
\ee
has a simple pole with residue $\alpha$ and is therefore best suited for an analysis
by Pad\'e approximants. We thus consider the series
\ba
D_C(1,u)&=&\frac{4}{u}\left(1+{\frac {20}{3}}{u}^{2}+{\frac {54791}{243}}{u}^{4}+{\frac {1879249}
{486}}{u}^{6}\right)\\
D_C(2,u)&=&\dfrac{8}{u}\left( 1-{\frac {45}{28}}{u}^{2}+{\frac {6445}{196}}{u}^{4}-{\frac {150331}{
92610}}{u}^{6}+{\frac {31831541863}{21003948}}{u}^{8}\right)
\label{dc}\\
D_C(3,u)&=&\dfrac{12}{u}\left(1-{\frac {91}{66}}{u}^{2}+{\frac
{4573}{242}}{u}^{4}-{\frac {2653298}
{59895}}{u}^{6}+{\frac {114561591157}{106732890}}{u}^{8}\right)\\
D_C(4,u)&=&\dfrac{16}{u}\left( 1-{\frac {51}{40}}{u}^{2}+{\frac {29791}{1800}}{u}^{4}-{\frac {
1262057}{27000}}{u}^{6}+{\frac {1055884297}{1215000}}{u}^{8}\right)
\ea
and model the full functions by Pad\'e approximants. 
Using the formulae
\be
C(u)=\exp{\int du\,D_C(u)},\quad
\frac{d}{du}f(u)=\int du \frac{C(u)}{u^2},
\label{conv}
\ee
the energy density can be reconstructed from the Pad\'e approximants to $D_C(u)$.
As an example, we show the $L+M=4,6,8$ approximants for the $N_t=2$ lattice in 
\fig\ref{series} (right).
Clearly, this sequence of approximants shows improved convergence compared to that
of the bare series.

\begin{table}
\vspace*{-1.5cm}
\begin{center}
\begin{tabular}{|c|c|c|c|}
\hline
 Pad\'e & Singularities   & Zeroes  & Residues
\\\hline\hline
$\left[0,6\right]$& $\pm\,1.8356$                   &                                 & $0.07486$ \\
                  & $\pm\,(0.2112\,\pm\,1.7192\,i)$ &                                 &           \\
$\left[2,4\right]$& $\pm\,1.6995$                   & $\pm\,2.8474$                   & $0.05656$ \\
                  & $\pm\,1.5674\,i$                &                                 &           \\
$\left[4,2\right]$& $\pm\,9.9475\,i$                & $\pm\,(1.1253\,\pm\,1.2410\,i)$ &           \\
\hline
$\left[0,8\right]$& $\pm\,1.6930$                   &                                 & $0.06080$ \\
                  & $\pm\,1.4349\,i$                &                                 &           \\
                  & $\pm\,(0.9152\,\pm\,1.6609\,i)$ &                                 &           \\
$\left[2,6\right]$& $\pm\,1.4893$                   & $\pm\,1.7142$                   & $0.02888$ \\
                  & $\pm\,1.5209\,i$                &                                 &           \\
                  & $\pm\,4.2104$                   &                                 &           \\
$\left[4,4\right]$& $\pm\,1.5502$                   & $\pm\,1.9430$                   & $0.03792$ \\
                  & $\pm\,1.4997\,i$                & $\pm\,2.2802\,i$                &           \\
$\left[6,2\right]$& $\pm\,0.1308\,i$                & $\pm\,0.1308\,i$                &           \\                         &                                 & $\pm\,(1.1278\,\pm\,1.2384\,i)$ &           \\
\hline
\end{tabular}
\begin{tabular}{|c|c|c|c|}
\hline
 Pad\'e & Singularities   & Zeroes  & Residues
\\\hline\hline
$\left[0,6\right]$& $\pm\,3.1022$                   &                                 & $0.13287$ \\
                  & $\pm\,1.6250\,i$                &                                 &           \\
$\left[2,4\right]$& $\pm\,3.0636$                   & $\pm\,3.0406$                   &  \\
                  & $\pm\,1.6228\,i$                &                                 &           \\
$\left[4,2\right]$& $\pm\,1.9628\,i$                & $\pm\,(0.9730\,\pm\,1.7187\,i)$ &           \\
\hline
$\left[0,8\right]$& $\pm\,2.0270$                   &                                 & $0.05936$ \\
                  & $\pm\,1.4583\,i$                &                                 &           \\
                  & $\pm\,(1.0967\,\pm\,1.5608\,i)$ &                                 &           \\
$\left[2,6\right]$& $\pm\,2.8687$                   & $\pm\,0.4335\,i$                & $0.11573$ \\
                  & $\pm\,0.4335i\,i$               &                                 &           \\
                  & $\pm\,1.6505i$                  &                                 &           \\
$\left[4,4\right]$& $\pm\,1.5084$                   & $\pm\,1.5762$                   &  \\
                  & $\pm\,1.4023\,i$                & $\pm\,1.5976\,i$                &           \\
$\left[6,2\right]$& $\pm\,0.8961\,i$                & $\pm\,0.8989\,i$                &           \\                         &                                 & $\pm\,(1.2985\,\pm\,1.5916\,i)$ &           \\
\hline
\end{tabular}
\end{center}
\caption{Singularities, zeroes and residues of $L+M=6,8$ Pad\'e approximants for 
$N_t=2$ (top) and $N_t=4$ (bottom). Residues are only given for those singularities
that enter the estimates for $\beta_c$, Table \ref{betac}.}
\label{sing}
\end{table}

The singularities in $D_C(u)$, indicated by zeroes of the denominator, 
together with the 
zeroes of the resulting approximants are shown in Table \ref{sing} for $N_t=2,4$, respectively.
Singularities in the immediate neighbourhood of a zero of the same approximant are typically
artefacts and unstable under variation of the approximant. 
However, for $N_t=2$ several 
approximants show poles around $\beta=1.5$ without zeroes in the immediate vicinity, indicating
that the full function indeed has a singluarity on the real axis in this region.
For $N_t=4$, on the other hand, the pole near $\beta=1.5$ is accompanied by a zero and not 
to be taken seriously. The next nearest pole on the real axis is instead around $\beta=2$.
A priori it is not possible to judge which approximants are better than others. 
The scatter in the results is thus a measure for the systematic error associated with 
the Pad\'e procedure. Moreover, there is also a scatter between approximants based
on different orders of the underlying series.
With increasing order, the approximants 
display more and more singularities which should eventually 
accumulate near the true singularity 
structure. 

To take these systematic effects into account, we estimate $\beta_c$ by averaging 
over the lowest lying real singularities obtained
from the two highest order approximants, i.e.~$L+M=4,6$ for $N_t=1$ and $L+M=6,8$ for $N_t=2-4$. 
To quantify the scatter due to the systematic uncertainties we quote 
$(\beta_c^{max}-\beta_c^{min})/2$ as an error estimate. The same procedure is followed
for the residues, and the results are collected in Table \ref{betac}.
Since the series are still short, the predictions for the critical coupling are not yet
very accurate, and those for the critical exponent even less so. Note, however,
that a first order phase transition has $\alpha=0$, whereas a second order
transition in the 3d Ising universality class has $\alpha=0.12$. Our results clearly
favour the latter, especially as the lattice becomes finer.

\begin{table}[b,t]
\begin{center}
\begin{tabular}{|c|c|c|c|}\hline
$N_t$   &  $\alpha$        &  $\beta_c$ & $\beta_c$ (Monte Carlo) \\ \hline\hline
1 & 0.061(38)  & 0.92(15)    & 0.85997(10)\cite{vel}  \\
2 & 0.052(19)  & 1.65(35)    & 1.880(3)\cite{fhk}     \\
3 & 0.078(50)  & 2.26(63)    & 2.177(3)\cite{fhk}     \\
4 & 0.102(37)  & 2.66(54)    & 2.299(6)\cite{fhk}     \\
\hline
\end{tabular}
\end{center}
\caption[]{Estimates for the critical coupling $\beta_c$ and the critical exponent 
of the deconfinement phase transition. The exponent for 3d Ising universality is 
$\alpha=0.12$.} 
\label{betac}
\end{table}

\subsection{Comparison with Monte Carlo data}

It is now interesting to compare the results from the strong coupling series with 
Monte Carlo simulations. The thermodynamic quantity most easily accessible by 
Monte Carlo is the energy density, which is simply the expectation value of the plaquette,
\be
e(\beta)=\frac{1}{6}\frac{d}{d\beta}f(\beta)
=\langle\,{\rm Tr}\,U_p\,\rangle_{N_t}-\langle\,{\rm Tr}\,U_p\,\rangle_{N_t=\infty}, 
\ee
where again the zero temperature (infinite $N_t$) piece is subtracted for renormalisation.
As we have seen, in the low beta region of the deconfined phase, the corresponding 
values are exponentially small, and very high statistics runs are necessary 
in order to get significant results for a quantitative comparison.
For the infinite volume vacuum lattice we have taken $12^4$, and $N_s=12, N_t=1,2,3,4$ for the 
finite $T$ lattices.
On the $N_t=2$ lattice up to $1.5\times 10^6$ field configurations were generated to achieve
sufficient accuracy, for the larger $N_t$'s this gets scaled down accordingly.

\begin{figure}[t!]
\hspace*{-0.7cm}
\includegraphics*[angle=-90,width=0.55\textwidth]{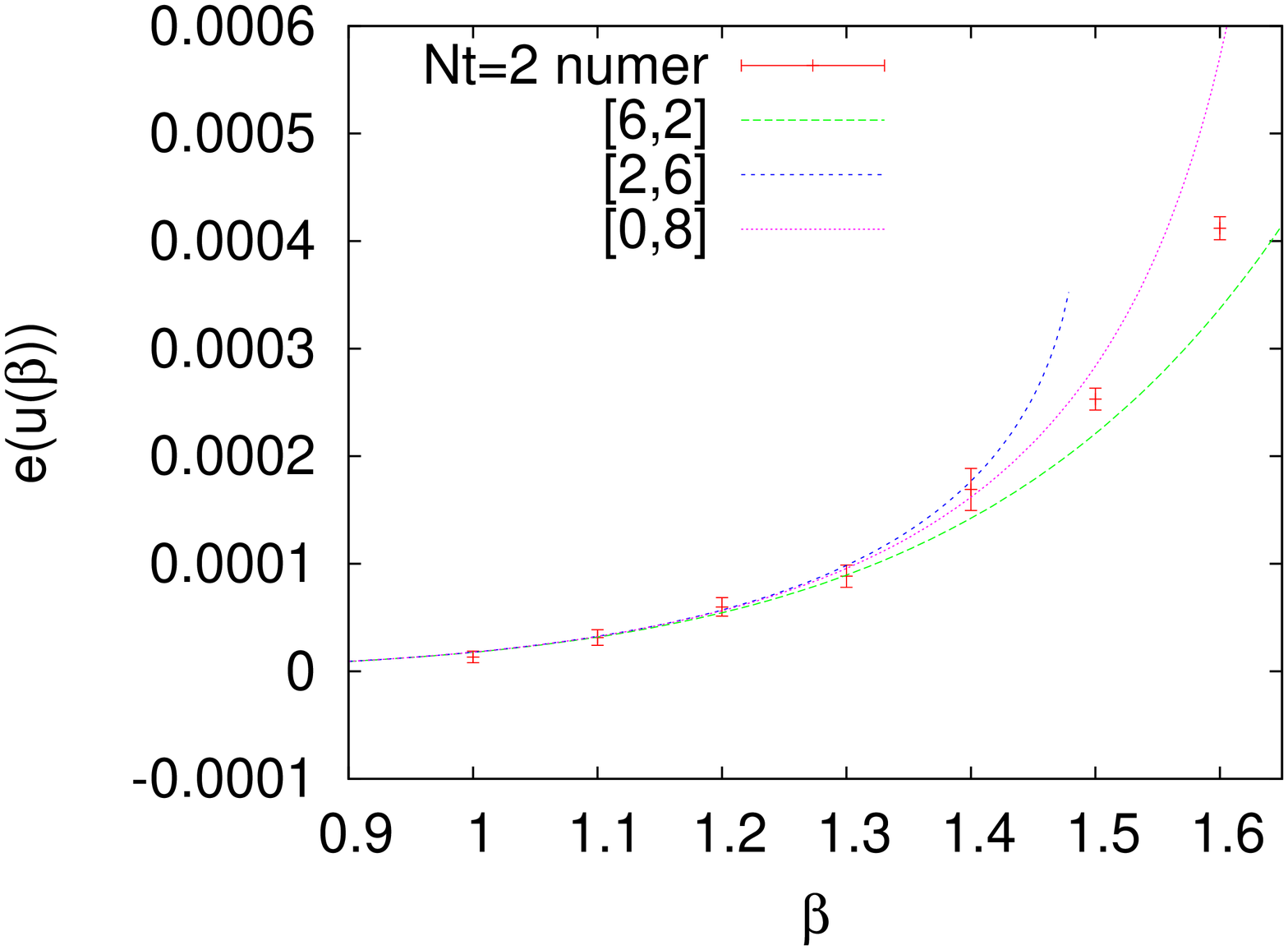}
\includegraphics*[angle=-90,width=0.55\textwidth]{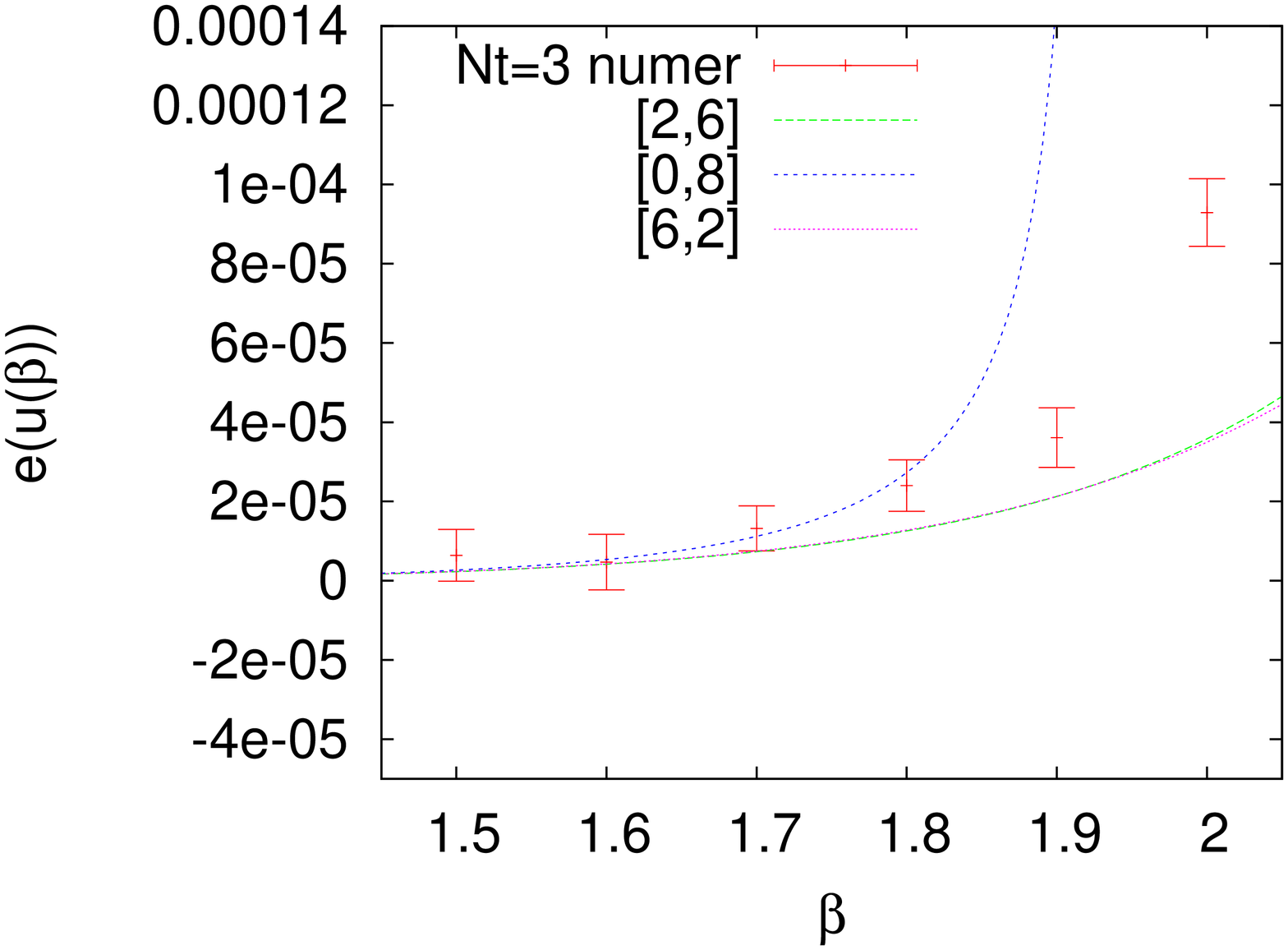}
\caption{Comparison of Monte Carlo data for $N_t=2$ (left) and $N_t=3$ (right) 
with Pad\'e approximants from the strong coupling series.
}
\label{ebeta}
\end{figure}

We compare these data with the best estimate based on the strong coupling series, i.e.~the
Pad\'e approximants to the highest available order in the logarithmic derivative of the 
heat capacity. 
Detailed results for $N_t=2,3$ are shown in \fig\ref{ebeta}. The different curves correspond to different
approximants to the same order, and thus serve as an error band quantifying the uncertainties
associated with the Pad\'e procedure, thus giving a valuable error estimate. 
We observe quantitative agreement with the lattice data all the 
way up to the lowest estimates of $\beta_c$. For $N_t=1,4$ we have checked at a few points that
a similar picture obtains. Thus, the error estimate based on our Pad\'e
analysis appears to be reliable and announces the breakdown of the validity of the series. 
\section{Screening masses \label{mass}}

Screening masses are defined by the exponential decay of the spatial correlation function
of suitable operators. An overview regarding definition, quantum numbers and numerical
results can be found in \cite{lp}. Here we consider the colour-electric field correlator 
$\langle \,\Tr F^a_{0i}(\bfx) \,\Tr F^a_{0i}(\bfy)\,\rangle$, which is in the $J^{PC}_T=0^{++}_+$
channel ($T$ denotes reflection in Euclidean time) containing the ground state and the 
mass gap. On the lattice, this corresponds to a correlation of temporal plaquettes, 
and the quantum numbers under the point group $D^4_h$ are $A_1^{++}$.

\begin{figure}
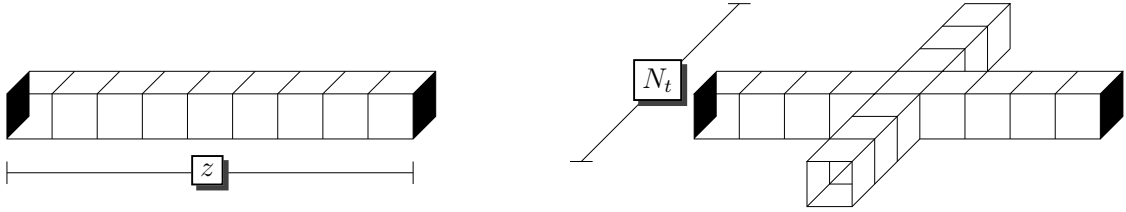

\begin{center}
\vspace*{2cm}
\begin{minipage}{7.5cm}
\hspace{2.5cm}
\scalebox{0.15}{
\pspolygon*[fillcolor=red](-18,7)(-18,3)(-16,5)(-16,9)
\pspolygon*[fillcolor=red](18,7)(18,3)(20,5)(20,9)
\pspolygon[linecolor=black](-18,7)(-18,3)(18,3)(20,5)(20,9)(-16,9)
\psline(-18,7)(18,7)(18,3)
\psline(18,7)(20,9)
\psline(-14,3)(-14,7)(-12,9)
\psline(-10,3)(-10,7)(-8,9)
\psline(-6,3)(-6,7)(-4,9)
\psline(-2,3)(-2,7)(0,9)
\psline(2,3)(2,7)(4,9)
\psline(6,3)(6,7)(8,9)
\psline(10,3)(10,7)(12,9)
\psline(14,3)(14,7)(16,9)
\psline(20,5)(18,5)
\psline(-18,0)(-1,0)\psline(-18,1)(-18,-1)
\psline(1,0)(18,0)\psline(18,1)(18,-1)\rput(0,0){\scalebox{6}{\psshadowbox{$z$}}}
}
\end{minipage}
\begin{minipage}{7.5cm}
\hspace{4cm}
\scalebox{0.15}{
\pspolygon*[fillcolor=red](-18,7)(-18,3)(-16,5)(-16,9)
\pspolygon*[fillcolor=red](18,7)(18,3)(20,5)(20,9)
\pspolygon[linecolor=black](-16,9)(0,9)(6,15)(10,15)(4,9)(20,9)(18,7)(2,7)(-4,1)(-8,1)(-2,7)(-18,7)
\psline(-18,7)(-18,3)(-6,3)
\psline(-8,1)(-8,-3)(-4,-3)(2,3)(18,3)(18,7)
\psline(18,3)(20,5)(20,9)
\psline(-4,-3)(-4,1)
\psline(2,3)(2,7)(4,9)
\psline(10,15)(10,11)(8,9)
\psline(-12,9)(-14,7)(-14,3)
\psline(-8,9)(-10,7)(-10,3)
\psline(-4,9)(-6,7)(-6,3)
\psline(4,9)(0,9)(-2,7)(2,7)
\psline(-6,3)(-2,3)(-2,-1)
\psline(-4,5)(0,5)(0,1)
\psline(2,11)(6,11)(6,9)
\psline(4,13)(8,13)(8,9)
\psline(6,3)(6,7)(8,9)
\psline(10,3)(10,7)(12,9)
\psline(14,3)(14,7)(16,9)
\psline(-8,-3)(-4,1)
\psline(-6,1)(-6,-1)(-4,-1)
\psline(-28,1)(-22,7)\psline(-29,1)(-27,1)
\psline(-20,9)(-14,15)\psline(-15,15)(-13,15)\rput(-21,8){\scalebox{6}{\psshadowbox{$N_t$}}}
}
\end{minipage}
\end{center}
\caption{Graphs contributing to the lowest order of the expansion of the 
screening mass at vanishing and finite temperature. 
The correlated plaquettes are black.}\label{scr_mass}
\end{figure}

Temporarily assigning separate gauge couplings to
all plaquettes, the correlator can be defined as \cite{sct0}
\begin{equation}
C(z)=\langle\mathrm{Tr}\,U_{p_1}(0)\,\,\mathrm{Tr}\,U_{p_2}(z)
\rangle=N^2\frac{\partial^2}{\partial\beta_1\partial\beta_2}\ln\,
Z(\beta,\beta_1\beta_2)\bigg\vert_{\beta_{1,2}=\beta}.
\end{equation}
At zero temperature the exponential decay is the same as for correlations in the 
time direction, and thus
determined by the glueball masses, the lowest of which may be extracted as
\begin{equation}
m=-\lim_{z\rightarrow\infty}\frac{1}{z}\ln\,C(z).\\
\end{equation}
The leading order graphs for the strong coupling series at zero temperature 
are shown in \fig\ref{scr_mass} (left).
This leads to the lowest order contribution:
\begin{equation}
C(z)=A\,u^{4z}=A\mathrm{e}^{-m_sz}.
\end{equation}
Thus, to leading order for the glueball mass is 
$m_s=-4\ln\,u(\beta)$.

Now we switch on a physical temperature, i.e.~keep the lattice volume compact in the
time direction. As in the case of the free energy, we are here only interested in
the temperature effects, i.e.~in the mass difference
\ba
\Delta m(T)=m(T)-m(0)&=&-\lim_{z\rightarrow \infty}\frac{1}{z}[\ln C(T;z)-\ln C(0;z)]\\
&=&-\lim_{z\rightarrow \infty}\left[\ln\left(1+\frac{\Delta C(T;z)}{C(0;z)}\right)\right],
\ea
with $\Delta C(T;z)=C(T;z)-C(0;z)$. A typical graph contributing in lowest order to this 
difference is shown in \fig\ref{scr_mass} (right). Summing up all leading and next-to-leading
order graphs gives
\be
\Delta m(T)=-\frac{2}{3}N_t\,u^{4N_t-6}\,c^{N_t}\,(1+4u^2),
\ee
i.e.~the screening masses decrease compared to their $T=0$ values.
As in the case of the free energy, due to the difference only $N_t$-dependent higher 
orders contribute to the temperature dependence of screening masses. 
Again, the leading order result is generic for all SU(N) and quantum number channels.
We conclude that in the confinement phase the lowest screening masses in each
quantum number channel should be close to the corresponding zero temperature particle masses,
with a significant temperature dependence showing up only near $T_c$.
This explains the findings of numerical investigations of
the lowest screening mass in SU(3) gauge theory, which for temperatures
as high as $T=0.97T_c$ see very little temperature dependence,
$\Delta m(T)/m(0)\gsim 0.83$ \cite{latmass}. 

\section{Conclusions \label{con}}

We have explored the possibilities of Euclidean strong coupling expansions for
finite temperature lattice Yang Mills theories in the confined phase. 
The general formalism applying to
zero temperature calculations can be taken over to this case, with the compact
time dimension affecting the type and number of graphs that contribute to a certain quantity.
As a consequence, temperature effects on the free energy density and screening masses
appear only at an $N_t$-dependent higher order and vanish exponentially as
$\beta=2N/g^2\rightarrow 0$. This explains the numerically observed exponential smallness
of the pressure and the near temperature independence of screening masses in the
confined phase as a typical strong coupling phenomenon.

We have explicitly calculated the first five terms of the series for the free energy density
in the case of SU(2). To the leading two orders, the result
agrees with that of an ideal glueball gas. This demonstrates that in the confined phase
the quasi-particles indeed correspond to the $T=0$ hadron excitations, as is typically assumed
in hadron resonance gas models. 
For lattices with $N_t=1-4$, we have analysed the series with
the help of Pad\'e approximants and estimated the critical couplings $\beta_c$
of the deconfinement phase transition. Since the series are still relatively short,
those results are not very accurate yet. However, within the estimated errors 
they are consistent with the critical couplings observed in Monte Carlo 
simulations. Moreover, up to the lowest estimated values of $\beta_c$ 
the Pad\'e approximants give a quantitative description of the Monte Carlo data for the 
equation of state.

In conclusion, the strong coupling expansion offers valuable qualitative insight
into temperature effects of the fully interacting, non-perturbative theory
in the confinement phase. We are currently extending this work
to the gauge group SU(3). 

\section*{Acknowledgement:}
We thank B.~Svetitsky for pointing us to the early literature, and Ph.~de Forcrand
for constructive comments on an earlier version of the manuscript.
This work is supported by the BMBF project
{\em Hot Nuclear Matter from Heavy Ion Collisions
     and its Understanding from QCD}, No.~06MS254.

\end{document}